\documentclass[prb,superscriptaddress,floatfix,twocolumn,showpacs,amsfonts,final]{revtex4}

\usepackage[utf8]{inputenc}
\usepackage[english]{babel}
\usepackage{graphicx}
\usepackage{color}
\usepackage{bm}
\usepackage{amssymb}
\usepackage[intlimits]{amsmath}

\begin{document}


\title{ Quantum Lifshitz criticality in a frustrated two-dimensional XY model}

\author{Yaroslav A. Kharkov}
\affiliation{School of Physics, University of New South Wales, Sydney 2052, Australia}
\affiliation{Joint Quantum Institute and Joint Center for Quantum Information and Computer Science, NIST/University of Maryland, College Park, Maryland 20742, USA}
\author{Jaan Oitmaa}
\author{Oleg P. Sushkov}
\affiliation{School of Physics, University of New South Wales, Sydney 2052, Australia}

\begin{abstract}
Antiferromagnetic  quantum spin systems  can exhibit a transition between 
collinear and  spiral ground states, driven by frustration.
Classically this is a smooth crossover and the crossover point is termed
a Lifshitz point.  
Quantum fluctuations change the nature of the transition.
In particular it has been argued previously that
in the two-dimensional (2D) case a spin liquid (SL) state is
developed in the vicinity of the Lifshitz point, termed a Lifshitz SL.
  In the present work, using a field theory approach, we solve the
  Lifshitz quantum phase transition  problem for the 2D frustrated XY-model.
  Specifically, we show that, unlike the SU(2) symmetric Lifshitz case, in the
  XY-model   the SL exists only at the critical point.
  At zero temperature we calculate nonuniversal critical exponents in the
  N\'eel and in the spin spiral state and relate these to properties of
  the SL.  We also solve the transition problem  at a finite temperature and
  discuss the role of topological excitations.
\end{abstract}
\maketitle

\section{Introduction.}
The Lifshitz point is the  classical (non-quantum) crossover between  a
collinear antiferromagnet  and a spin spiral state. The crossover  is
driven by frustration. Some time ago Ioffe and Larkin pointed out that if a frustrated 2D antiferromagnet is tuned to the Lifshitz point,
 long wavelength quantum fluctuations destroy the long-range order and lead to the formation of a quantum disordered phase, a spin
liquid (SL).~\cite{Ioffe1988} In the present paper we term this phase a Lifshitz SL. It has properties which are different
from other known types of SL. Locally it maintains antiferromagnetic or
spin spiral correlations. For this reason quantum field theory is
the most natural technique to describe the state, and this technique
was used already in the pioneering work~\cite{Ioffe1988}.
After the first work the idea   of Lifshitz quantum criticality and
Lifshitz SL was almost forgotten.
However recently it has attracted more attention.
Lifshitz quantum criticality arises in frustrated
quantum magnets with competing interactions~\cite{Ardonne2004,Capriotti2004,
Fradkin2013,Balents2016,Kharkov2018} and in
underdoped cuprates where the effective frustration is due to itinerant
holes~\cite{Kharkov2018a,Kharkov2018b}.
Interestingly, the XY-type Lifshitz field theories arise also in
Rokhsar-Kivelson dimer models\cite{Rokshar1988, Ardonne2004}, as well as
in liquid crystals \cite{Chaikin1995},
Bose Einstein condensates of ultracold atoms~\cite{Po2015},
and even in some cosmological
models~\cite{Horava2009, Hoyos2014}.
A Lifshitz point in a classical  frustrated XY model at a finite temperature was also considered in Ref.~\cite{Schenck2014}.

In the present work we study  2D frustrated magnets with competing interactions
and solve  the problem of a Lifshitz quantum phase transition  in
the 2D frustrated XY-model.
We derive a field theoretical description for the $O(2)$ symmetric model in the vicinity of the Lifshitz transition. This allows us to calculate nonunversal  critical exponents
in the N\'eel and in the spin spiral state and  relate these to properties of the SL.    We also study the Lifshitz transition
at a finite temperature, accounting for both quantum
and thermal fluctuations. We discuss the
relative importance of
the perturbative and the topological (vortex) excitations.
We also underline differences
  between Lifshitz criticalities  for the XY and the SU(2)-symmetric case.

  The paper is organised as follows. In Sec. \ref{series}
  we discuss the $J_1-J_3$ Heisenberg model on the square lattice, $S=1/2$, as an example
  of Lifshitz criticality. Here we present results of  numerical
  series expansion calculations that motivate further analytical analysis.
  Section \ref{zeroT} presents  analytical solution of the problem
  at zero temperature.
  Finite temperature properties are considered in Section \ref{finiteT}.
Section \ref{sec:concl}  summarizes our conclusions.

\section{$J_1-J_3$ model, series expansions}  \label{series}
Before presenting the general solution in the framework of a quantum field
theory, we consider a specific lattice model. One of the simplest 2D spin systems with frustration induced by competing interactions is   
 $J_1-J_3$ Heisenberg antiferromagnet on the square lattice,
\begin{equation}
  H = J_1 \sum_{{\alpha}\langle ij \rangle}  S_i^{\alpha} S_j^{\alpha}
  + J_3 \sum_{{\alpha}\langle \langle \langle   ij \rangle \rangle \rangle}
  S_i^{\alpha} S_j^{\alpha},
  \label{eq:H}
\end{equation}
where $\bm S_i$ is spin 1/2 at the lattice site $i$. Antiferromagnetic
interactions, $J_{1,3}>0$, account for the nearest neighbor sites, and the third
neighbor sites.
 If the summation over $\alpha$ is performed over
all components of spin, $\alpha=x,y,z$, the model is SU(2)-symmetric,
or O(3)-symmetric.
The case $\alpha=x,y$ corresponds to the XY-model, with $O(2)$ symmetry.

Frustrated $J_1-J_2$ and $J_1-J_2-J_3$ models have been discussed in
numerous studies, see e.g. Refs.~\cite{Ferrer93,Sushkov2001,Capriotti04,
Sindzingre2010,Reuther11a,Zhu13,Bishop15}.
In the $J_1-J_2$ model the nontrivial regime is realised around
$J_2/J_1 \sim 0.5$. In this regime energies of the  N\'eel state,
the spin spiral state, and the spin stripe state are close.
The spin stripe state ``spoils'' the situation. It has
very different short-range spin structure and this structure is strongly
mixed up by quantum fluctuations.
Therefore the $J_1-J_2$ model definitely does not belong to the Lifshitz class.
However in $J_1-J_3$ model the situation is different. In the classical limit,
$S \gg 1$, there is a Lifshitz point at $J_3=J_1/4$ with a smooth formation
of the spin spiral at $J_3>J_1/4$.
The spin stripe phase with the different short-range order has a significantly
higher energy, and does not play a role at low temperature.
So the $J_1-J_3$ model likely belongs to the Lifshitz class.

\begin{figure}[h!]
\includegraphics[scale=0.13]{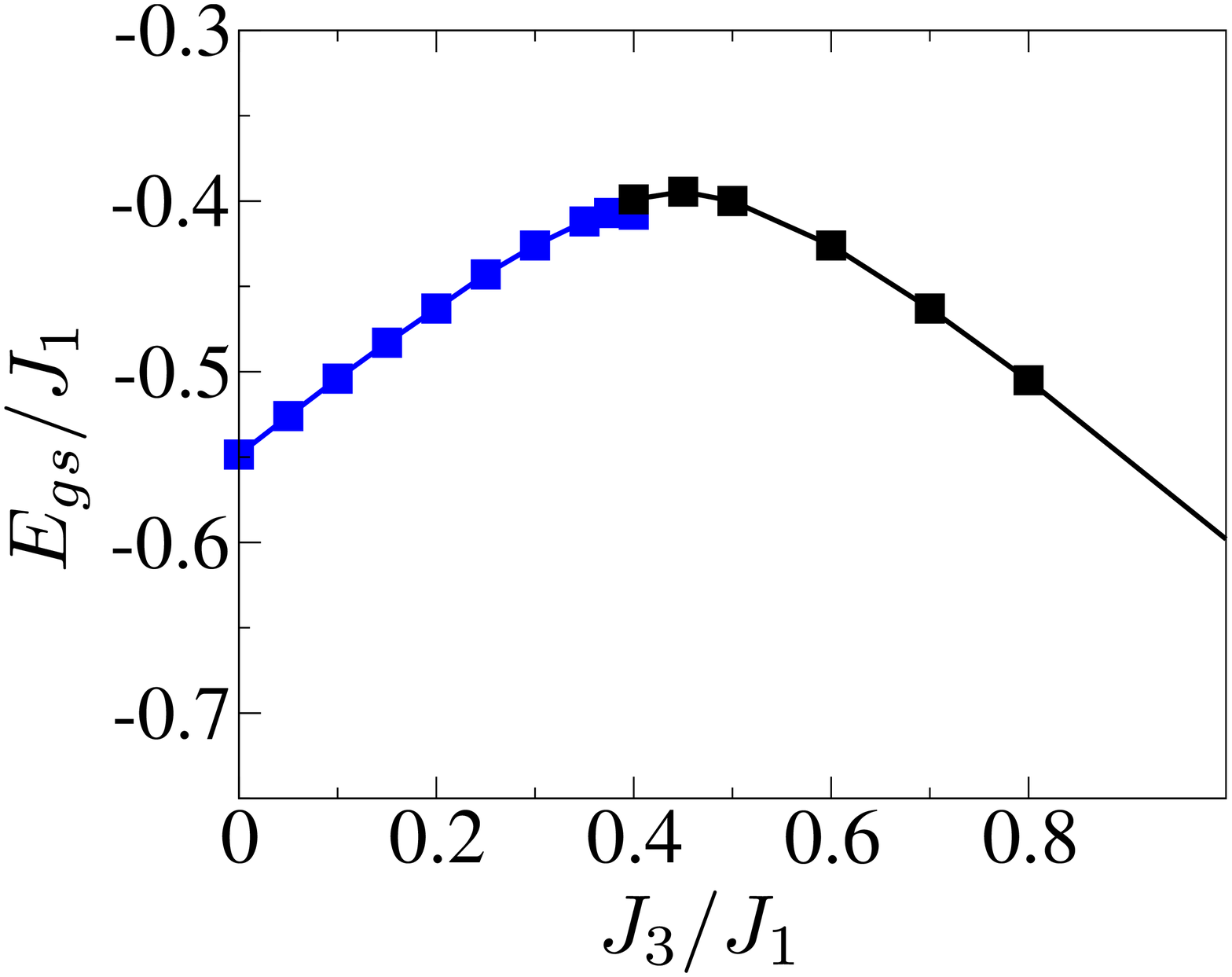}
\includegraphics[scale=0.13]{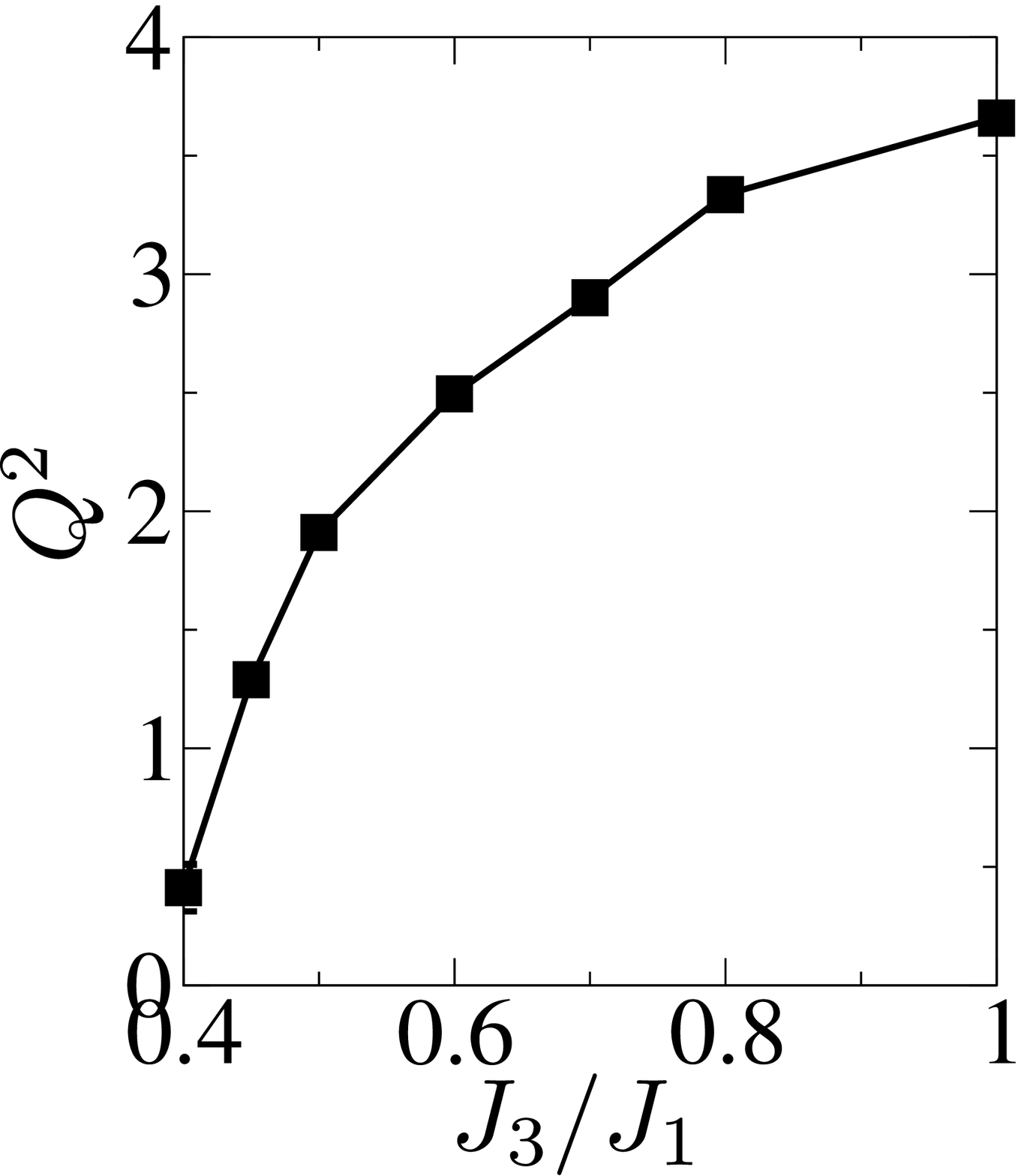}
\includegraphics[scale=0.13]{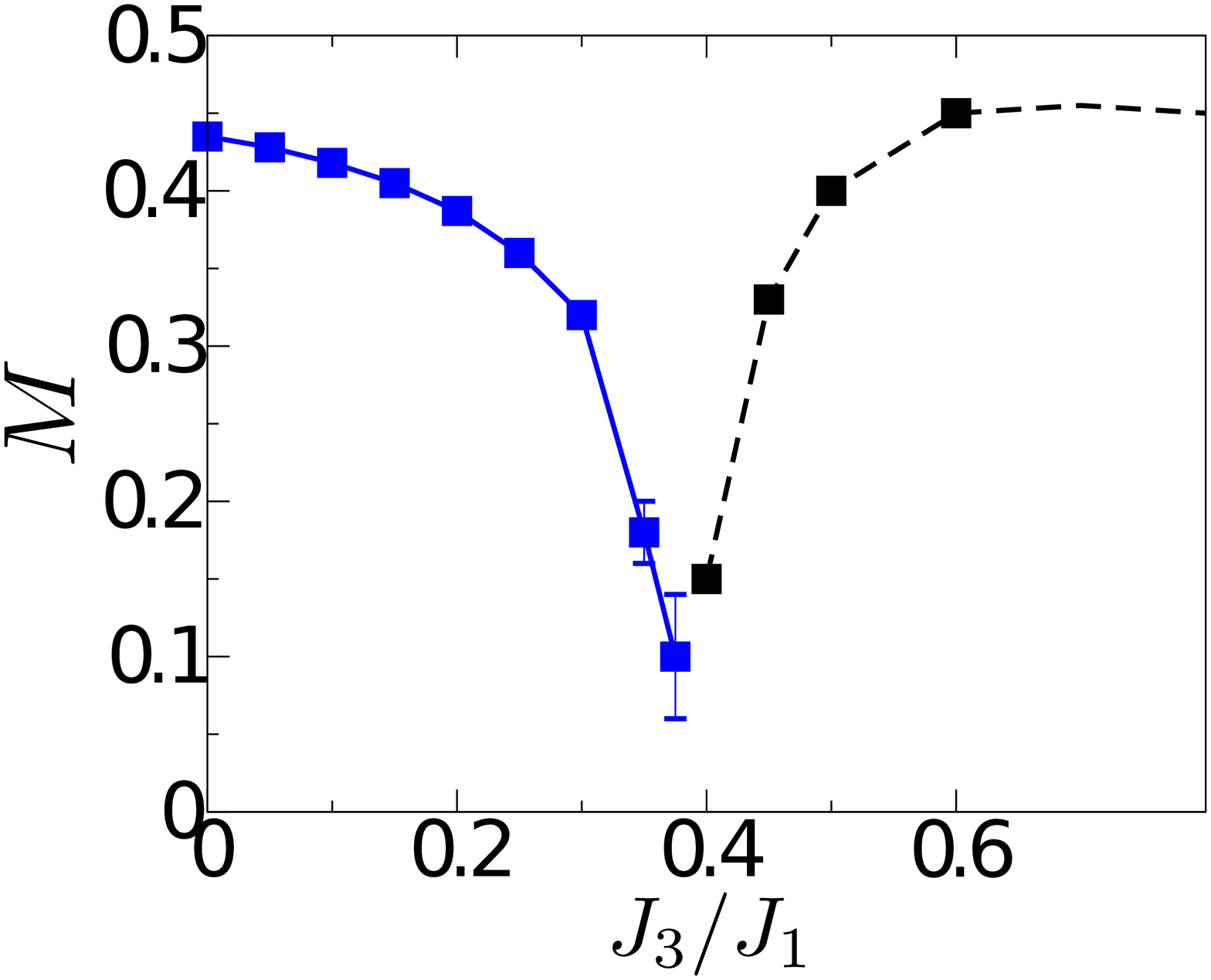}
\caption{The $O(2)$ $J_1-J_3$ model:
  At $J_3/J_1 < 0.4$ the ground state is the  N\'eel state,
  and at $J_3/J_1 > 0.4$ the ground state is the  spin spiral state.
  The spin liquid is realised at one point, $J_3/J_1 \approx 0.4$.
    Panel a: Energy of the ground state versus $J_3/J_1$.
  Panel b: The spin spiral wave vector squared versus $J_3/J_1$.
  Panel c: Static magnetisation $M$ versus $J_3/J_1$.
}
\label{fig:O2}
\end{figure}

\begin{figure}[h!]
\includegraphics[scale=0.13]{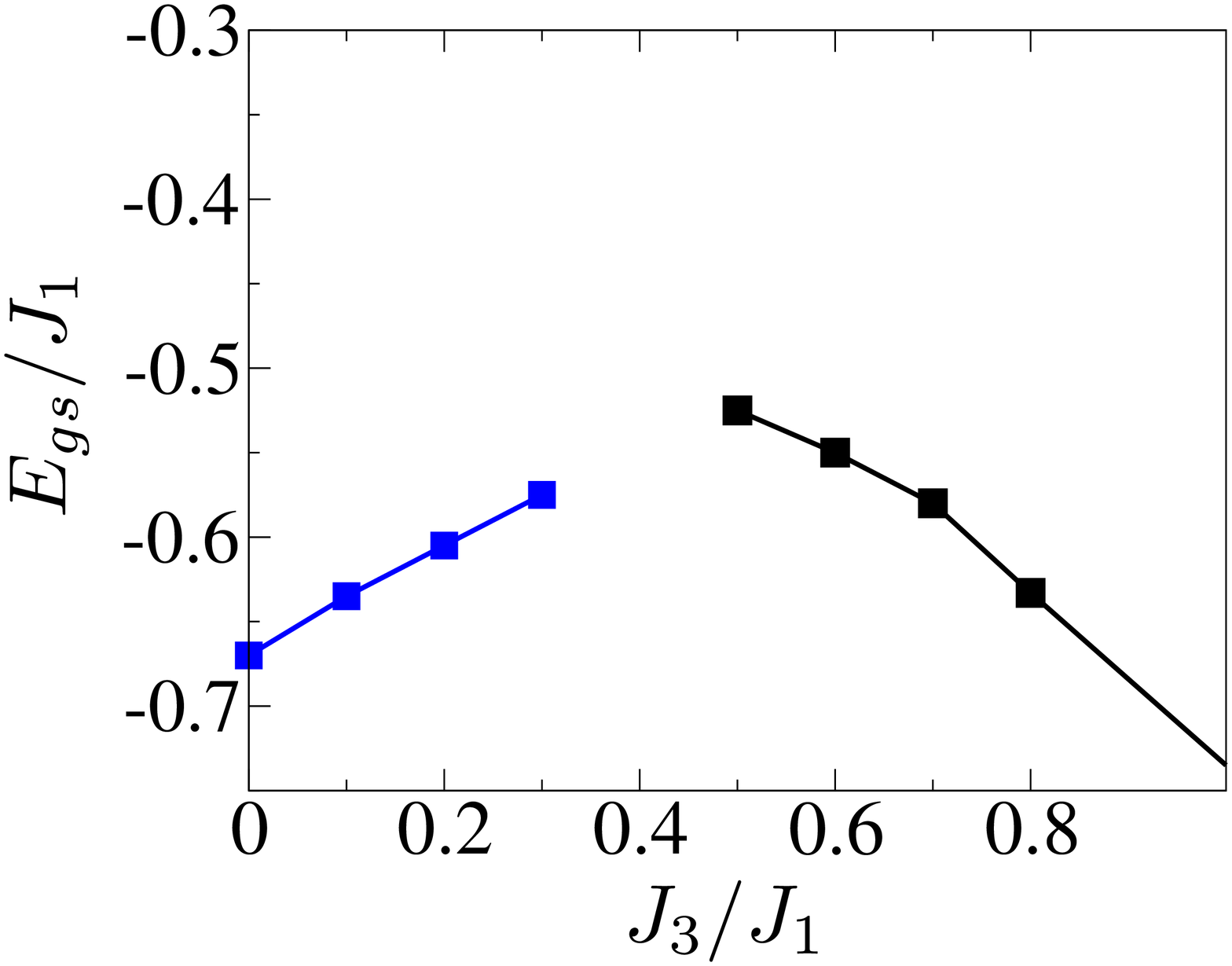}
\includegraphics[scale=0.13]{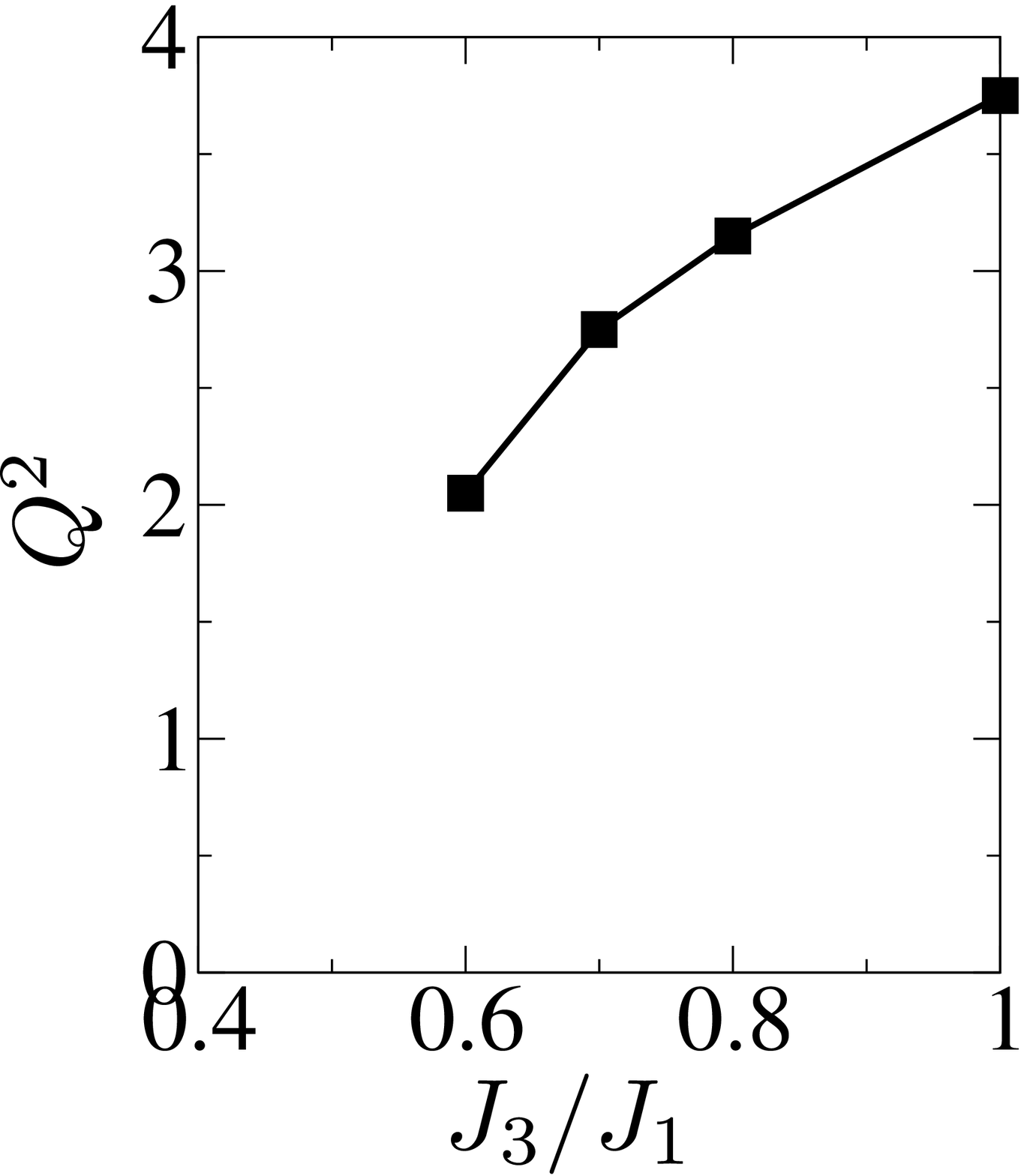}
\includegraphics[scale=0.13]{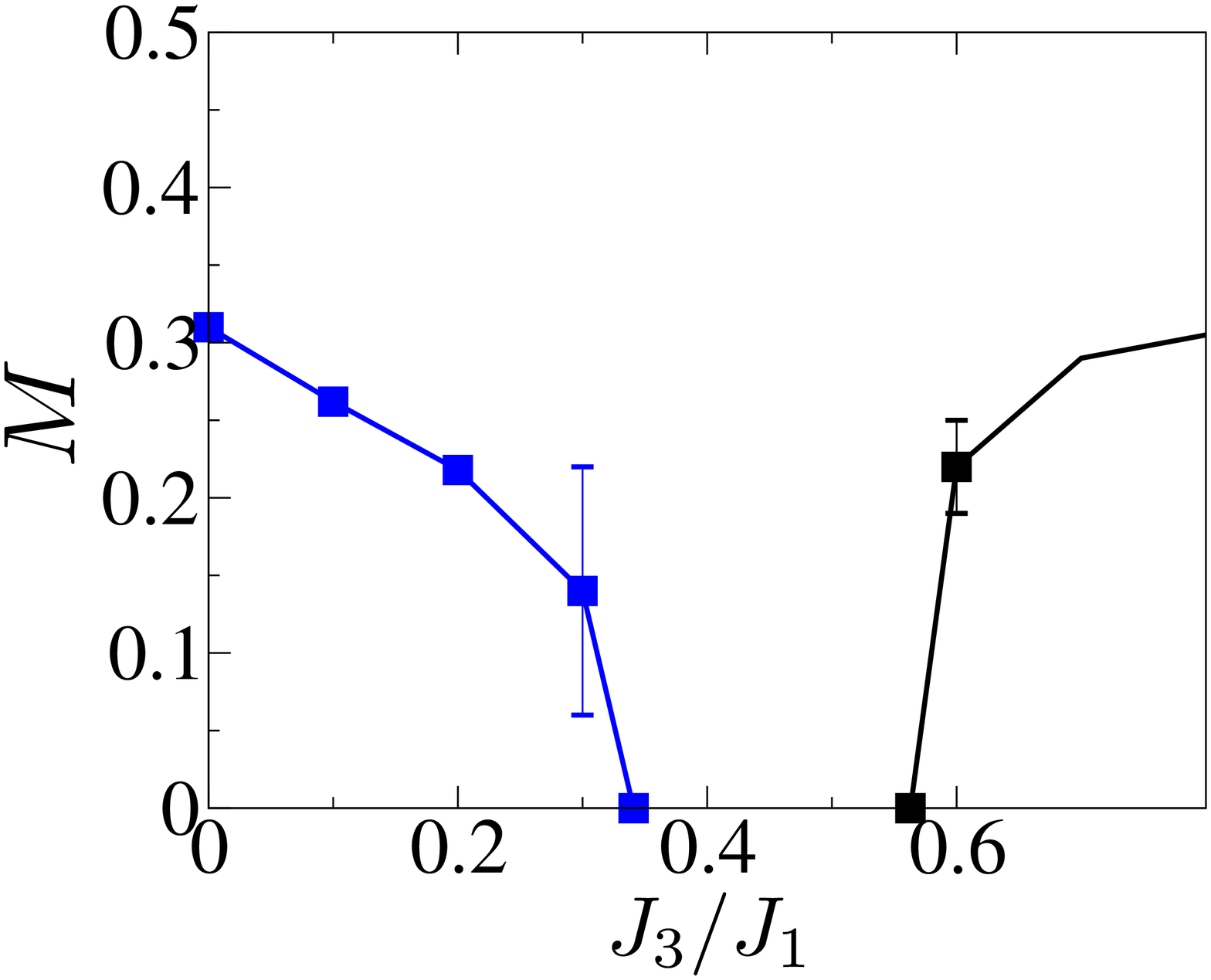}
\caption{The $O(3)$ $J_1-J_3$ model, results of Ref.~\cite{Kharkov2018}
At $J_3/J_1 < 0.33$ the ground state is the  N\'eel state,
  and at $J_3/J_1 > 0.55$ the ground state is the  spin spiral state.
  The quantum disordered (spin liquid) phase is realised within the range $0.33 <J_3/J_1 < 0.55$.
  Panel a: Energy of the ground state versus $J_3/J_1$.
  Panel b: The spin spiral wave vector squared versus $J_3/J_1$.
  Panel c: Static magnetisation versus $J_3/J_1$.
  }
\label{fig:O3}
\end{figure}

We have performed extensive series expansion calculations\cite{OitmaaBook} both in the N\'eel phase  and
the spin-spiral phase.
Unfortunately the series expansion method is not able to probe properties
of the  SL phase directly.
However, the method allows us to find the range of parameters where the SL exists.
In the N\'eel phase the series starts from the simple Ising antiferomagnetic
state.
In the spin spiral phase the calculation is more tricky. We first impose a
classical diagonal spiral with
some  wave vector $Q$ and find the total energy of this state $E(Q)$.
This includes the classical energy and the
quantum corrections calculated by means of series expansions. We perform
this calculations for many
values of $Q$ and then find numerically the minimum of $E(Q)$. Such procedure
gives us the
ground state energy $E_{gs}$ and the physical wave vector $Q$.
Plots of the ground state energy $E_{gs}$, the spiral vector Q, and the static
on-site magnetization M for the XY $J_1-J_3$ model are presented in
Fig.~\ref{fig:O2}.
For comparison in Fig.~\ref{fig:O3} we present the same quantities
calculated in Ref.~\cite{Kharkov2018} for the $O(3)$ $J_1-J_3$ model.

Note that according to our series calculations the period of the
incommensurate spin spiral is quite long.
For example $Q^2=2$ implies that $Q_x=Q_y=1=1/(2\pi)$r.l.u. So, the period in
each direction is $2\pi \approx 6$ lattice spacing.
While the series expansion method does not have a problem with this,
it can be quite hard to obtain such a long period in a finite cluster calculation. 
Moreover, the ground state is 4-fold degenerate,
$\bm Q= \frac{1}{\sqrt{2}}(\pm Q,\pm Q)$. This is on the top of the
continuous O(N) degeneracy that results in a well understood
rotational tower of quantum states in a finite
system~\cite{Anderson1952,Misguich2002}.
In the spiral case the continuous group tower must consist of 4-fold
split (or degenerate) states due to the $Q-$degeneracy, and the splitting depends
on the geometry of the cluster. It could be nontrivial to disentangle
such a spectrum in a finite size cluster numerically.

Concluding this section,
the most important qualitative difference between $O(3)$ and $O(2)$
is the size of the ``window'' in the parameter space occupied by the SL phase.
In the $O(3)$ case this is finite interval and in the $O(2)$ case it is likely that
the interval shrinks to a point.
Remarkably the problem of the critical behaviour of the $O(2)$ Lifshitz model
can be solved exactly. Now we proceed to the solution.

\section{Zero temperature solution} \label{zeroT}
In the vicinity of the Lifshitz point the spin dynamics is described by
an effective nonlinear $\sigma$ model with the Lagrangian~\cite{Ioffe1988}
\begin{eqnarray}
  \label{eq:L}
\mathcal{L} = \frac{\chi_\perp}{2}(\partial_t n_\mu)^2 - \frac{1}{2}n_\mu K(\partial_i) n_\mu \ .
\end{eqnarray}
Here $n_\mu$ is the vector of staggered magnetisation, with ${\bm n}^2=1$,
$\partial_i$ are the spatial gradients $(i=x,y)$,
$\chi_\perp$ is the transverse
magnetic susceptibility. 
The general form of the ``elastic energy'' operator $K(\partial_i)$  assuming
that the $n$-field is sufficiently smooth is
\begin{equation}\label{eq:K(q)}
K(\partial_i) = -\rho(\partial_i)^2 + b_1(\partial_x^4+\partial_y^4)+2b_2\partial_x^2 \partial_y^2 + \mathcal{O}(\partial_i^6) .
\end{equation}
The kinematic form of the Lagrangian (\ref{eq:L}) is dictated by global
symmetries of the system.
In the $O(3)$ case the vector ${\bm n}$ has three components and in the
$O(2)$ case two components. Hereafter we consider $O(2)$ and hence the vector can
be parameterised by the single angle $\theta$
\begin{eqnarray}
  \label{tetp}
\bm n = (\cos{\theta}, \sin{\theta}).
\end{eqnarray}
The spin stiffness $\rho$ is the tuning parameter that drives the system
across the Lifshitz transition. 
The spin stiffness is positive in the N\'eel phase, negative  in the spiral
phase  and vanishes  at the Lifshitz point. The 4th
order spatial derivative $b$-terms
are necessary for stabilisation of the spiral order at negative $\rho$,
and we assume  $b_{1,2}>0$.

The Lagrangian (\ref{eq:L}) can be rewritten in terms of the angle $\theta$:
\begin{eqnarray}
  \label{eq:L_full_theta}
&&\mathcal{L} = \chi_\perp\frac{(\partial_t\theta)^2}{2} - \frac{\rho (\partial_i \theta)^2}{2} \nonumber \\
&&- \frac{b_1}{2} \left[ (\partial_x^2 \theta)^2 + (\partial_y^2 \theta)^2 + (\partial_x \theta)^4 + (\partial_y \theta)^4\right] 
\nonumber\\
&& - b_2 \left[ (\partial^2_{xy} \theta)^2 + (\partial_x\theta)^2 (\partial_y\theta)^2 \right].
\end{eqnarray}
Classically the ground state of this Lagrangian is the collinear N\'eel state
at $\rho > 0$, $\theta=const$ and the ground state is  the spin spiral at
$\rho<0$, $\theta = \bm {Qr}$,  where $\bm Q$ is the wave vector of
the spiral. 
For $b_1\leq b_2$ the spiral wave vector is directed along $x$ or $y$,
and for  $b_1>b_2$ the wave vector is directed along the main diagonals.
\begin{eqnarray}
  \label{Qb}
  b_1\leq b_2: \ \ \  && \bm Q= (\pm Q,0), (0,\pm Q)\nonumber\\
                      && Q^2=|\rho|/(2b_1) \nonumber\\
  b_1 >  b_2: \ \ \   && \bm Q= \frac{1}{\sqrt{2}}(\pm Q,\pm Q), \ \ \
  \frac{1}{\sqrt{2}}(\pm Q,\mp Q)\nonumber\\
                      &&Q^2 = |\rho|/(b_1+b_2)
\end{eqnarray}

Parameters of the effective Lagrangians (\ref{eq:L}),(\ref{eq:L_full_theta})
can be expressed in terms of the parameters of the Heisenberg model.
These are expansions in powers of spin $S$. For the $J_1-J_3$ Heisenberg model
on the square lattice in the leading order in $S$
the parameters are~\cite{Ioffe1988, Kharkov2018}
\begin{eqnarray}
\label{parS}      
&&\rho= S^2(J_1-4J_3)\nonumber\\
&&b_1=\frac{S^2}{24}(-J_1+16 J_3)\nonumber\\
&&b_2=0\nonumber\\
&&\chi_\perp = 1/(8J_1)
\end{eqnarray}

\begin{figure}
\includegraphics[scale=0.4]{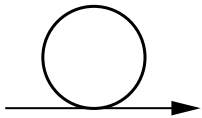}
\caption{ Single loop self-energy  diagram for  $\theta$-field generated by the quartic term in  Eq. \ref{eq:L_full_theta}.
  }
  \label{fig:loop}
\end{figure}

The Lagrangian (\ref{eq:L_full_theta}) contains quartic in $\theta$ terms.
These terms lead to the self-energy of the field $\theta$. The single
loop self-energy diagram is shown in Fig.~\ref{fig:loop}.
The self-energy has an ultraviolet part that is not singular at $\rho \to 0$
and it also has an infrared part singular at $\rho \to 0$.
The ultraviolet part just gives subleading corrections in S to equations
(\ref{parS}). 
For example, the quartic terms result in  the positive ultraviolet correction
to $\rho$  which extends the region of the Neel phase in  Fig.\ref{fig:O2}. The Lifshiz point shifts to
$J_3/J_1\approx 0.4$ from the classical Lifshits point $J_3/J_1=0.25$ that follows from Eqs. (\ref{parS}).
It is interesting to note that quantum fluctuations extend the
  region of the N\'eel phase, this is the general property of
  Lifshitz criticality.~\cite{Kharkov2018}
 On the other hand in the $J_1-J_2$ model
  quantum fluctuations shrink the region of the N\'eel phase.
  The classical critical point is $J_2/J_1=0.5$ and the quantum
  critical point is $J_2/J_1=0.38$.~\cite{Sushkov2001}
  This observation confirms our point that the $J_1-J_2$ model does not
  belong to the Lifshitz class.

The infrared part of the self energy is more interesting, as it influences
critical properties, and we discuss this point below.
The transverse susceptibility $\chi_\perp$ does not have an infrared
divergent correction, and in the single loop approximation,
Fig.~\ref{fig:loop}, the correction is actually zero. Therefore
hereafter we set $\chi_\perp = 1$ and hence use dimensionless parameters
\begin{eqnarray}
  \label{dim}
  && \rho \to \chi_\perp \rho\nonumber\\
  &&b_{1,2}\to \chi_\perp b_{1,2}\nonumber\\
  && T \to \chi_\perp T
  \end{eqnarray}
The last line defines a dimensionless temperature that we use in the next section.

The values of the coefficients $b_1$ and $b_2$ depend on the specific choice of
the lattice model. 
The special case $b_1=b_2$ corresponds to the situation when the ground
state energy is degenerate with respect to the direction of $\bm Q$.
In this degenerate case one has to account for 
higher order terms  in the powers of the spatial gradients $\mathcal{O}(\partial_i^6)$ in the expansion of the elastic energy $K(\partial_i)$.
In this work we assume that the system is far away from the degeneracy point,
$b_1=b_2$.

\subsection{Spin stiffness renormalization}
Let us approach the Lifshitz point from the N\'eel phase, $\rho > 0$.
Quadratic in $\theta$ terms of the Lagrangian (\ref{eq:L_full_theta})
result in  the following dispersion
\begin{eqnarray}
\label{dispq}
\omega_q = \sqrt{\rho q^2 + b_1 (q_x^4 + q_y^4) + 2 b_2 q_x^2 q_y^2} \ ,
\end{eqnarray}
where $q$ is the momentum.
At the Lifshitz point, $\rho=0$, the dispersion is quadratic in momentum,
and the dynamical critical exponent is $z=2$,
\begin{eqnarray}
\label{disp0}
\omega_{0q} = \sqrt{b_1 (q_x^4 + q_y^4) + 2 b_2 q_x^2 q_y^2} \ .
\end{eqnarray}
To calculate the self-energy Fig.~\ref{fig:loop} we decouple 
the nonlinear terms in the Lagrangian (\ref{eq:L_full_theta})
\begin{eqnarray}
  (\partial_i \theta)^4  \rightarrow 6 (\partial_i \theta )^2  \langle  (\partial_i \theta )^2 \rangle =
  3  (\partial_i \theta )^2 \int \frac{d^2 q}{(2\pi)^2} \frac{q_i^2}{\omega_q } .
\end{eqnarray}
Thus, the self-energy results in a correction to the spin stiffness.
First we perform the ultraviolet renormalization. To do so consider the
Lifshitz point, $\rho=0$. Here the correction is
\begin{eqnarray}
  \label{dro}
  \Delta \rho= \left(3 b_1+\frac{b_2}{2}\right)\int \frac{d^2 q}{(2\pi)^2}
  \frac{q_x^2}{\omega_{0q} } \ .
\end{eqnarray}
The integral is  convergent  in the infrared limit $(q\to 0)$ in spite of the quadratic dispersion.
Hence, $\Delta \rho$ is just a constant that has to be added  to the relation (\ref{parS}).
We absorb this shift in the definition of $\rho$.
Away from the Lifshitz point in the N\'eel phase the self-energy
leads to a logarithmic renormalization of the spin stiffness,
$\rho \to \rho_r$.
Similar to (\ref{dro}) we find
\begin{eqnarray}
  \label{eq:rho_r_sub1}
&&\rho_r = \rho +  
 \left(3b_1+\frac{b_2}{2}\right)
  \int \frac{d^2 q}{(2\pi)^2}  q_x^2 \left[\frac{1}{\omega_q} -
    \frac{1}{\omega_{0q}} \right] .
\end{eqnarray}
For self consistency in the dispersion $\omega_q$  in this formula
we must replace $\rho \to \rho_r$, see Eq.(\ref{dispq}). Evaluation of the
integral in (\ref{eq:rho_r_sub1}) with logarithmic accuracy is straightforward.
This leads to the following equation for the renormalised spin stiffness
$\rho_r$
\begin{equation}
  \label{rcor}
  \rho_r = \frac{\rho }{1 + \frac{3\kappa_0 }{8\pi \sqrt{b_1}}
\left(1+\frac{b_2}{6b_1}\right)
    \ln\left(\frac{\Lambda^2 b_1}{|\rho_r|}\right) } \ .
\end{equation}
Here $\Lambda \sim \pi/a$ is the ultraviolet momentum cutoff,
$a$ is the lattice spacing, and the number $\kappa$ is
the following angular integral 
\begin{eqnarray}
  \label{kapn}
  \kappa_0 = \int_0^{2\pi} \frac{d\alpha/(2\pi) \ \cos^2{\alpha}}
{[(\cos^4{\alpha} + \sin^4{\alpha} ) + 2(b_2/ b_1) \sin^2{\alpha} \cos^2{\alpha}]^{3/2}} 
\end{eqnarray}

The logarithmic correction to the spin stiffness can not depend
on the sign of $\rho$, so (\ref{rcor})  is valid also for $\rho<0$.
This point is obvious since the logarithmic correction comes from
quantum fluctuations in the  momentum range $Q\sim \sqrt{|\rho|/b_1} < q < \Lambda$.
This range is insensitive to formation of the spin spiral. 
This implies  that in Eqs. (\ref{Qb}) one must replace $\rho \to \rho_r$
and the correct scaling of the wave vector is
\begin{eqnarray}
  \label{Qb1}
  Q \propto \sqrt{\rho_r} \propto \sqrt{\frac{J_3-J_{3c}}
    {1+A\ln\left(\frac{B}{J_3-J_{3c}}\right)}} \ ,
  \end{eqnarray}
where the constants A and B are given by Eq.(\ref{rcor}).
So, there is a logarithmic correction to the power scaling of Q.
Besides the general scaling arguments one can check the relation
$Q \propto \sqrt{\rho_r}$ by a detailed calculation.
To do so we substitute $\theta = \bm Q \bm r +\phi$ in the Lagrangian
(\ref{eq:L_full_theta}) and expand it up to cubic terms $(\partial_i\phi)^3$.
A straightforward decoupling procedure
$(\partial_i\phi)^3\to 3(\partial_i\phi ) \langle(\partial_i\phi)^2\rangle$ leads directly to Eq. (\ref{rcor}).
The logarithmic correction in Eq. (\ref{Qb1}) is small. Therefore
the deviation of Fig.\ref{fig:O2}b from a straight line is due to the higher
powers of $J_3-J_{3c}$ that cannot be captured by the field theory. 

It is interesting to note that nonlinear terms in the Lagrangian (\ref{eq:L_full_theta}) do not renormalize the higher derivative terms $b_{1,2}(\partial^2\theta)^2$. Such corrections can only be generated by subleading terms, e.g. $(\partial^2\theta)^4$ or $(\partial^2 \theta)^2(\partial\theta)^2$.

In the spin spiral phase it is convenient to describe excitations in terms of the  field $\phi$.
The substitution $\theta = \bm Q \bm r +\phi$ in the Lagrangian (\ref{eq:L_full_theta}) 
results in the following quadratic form
\begin{equation}\label{eq:E_2_spir}
 \frac{1}{2}\widetilde\rho_{ij} \partial_i  \phi \partial_j \phi + \frac{b_1}{2}\left[(\partial_{xx} \phi)^2+(\partial_{yy} \phi)^2\right] + b_2(\partial_{xy}\phi)^2 .
\end{equation}
Hence the effective spin stiffness tensor $\widetilde \rho_{ij}$ that
determines the spectrum of excitations in the spin spiral phase reads:
\begin{eqnarray}\label{eq:rho_ij}
&&\widetilde\rho_{ij}=
2|\rho_r|
\begin{pmatrix}
&1 &0\\
&0 &\frac{b_2}{b_1}-1
\end{pmatrix}, \quad b_1\leq b_2; \nonumber
\\
&&\widetilde\rho_{ij}=
2|\rho_r|
\begin{pmatrix}
&\frac{b_1}{b_1+b_2} &\frac{b_2}{b_1+b_2}\\
&\frac{b_2}{b_1+b_2} &\frac{b_1}{b_1+b_2}
\end{pmatrix}, \quad b_1> b_2.
\end{eqnarray}
Hence the excitation spectrum in the spin spiral phase is
\begin{eqnarray}
\label{dispq1}
\widetilde{\omega}_q
= \sqrt{\widetilde\rho_{ij}q_iq_j + b_1 (q_x^4 + q_y^4) + 2 b_2 q_x^2 q_y^2} \ .
\end{eqnarray}

\subsection{Spin-spin correlator}
From Eq.(\ref{dispq}) we conclude that the phase fluctuation
in the N\'eel phase reads
\begin{eqnarray}
  \label{t2}
\langle \theta^2(r)\rangle - \langle \theta(r)\rangle^2 =\int\frac{d^2q}{(2\pi)^2}\frac{1}{2\omega_q} .
\end{eqnarray}
The fluctuation remains finite everywhere except of the point $\rho=0$
where the integral in (\ref{t2}) is infrared divergent.
This confirms the hint given by Fig.\ref{fig:O2}c that the spin liquid is
realised only at one point. Everywhere else the long range order is preserved.
To calculate the spin-spin correlator in the SL phase we follow the method of
Ref.\cite{Ardonne2004}
The equal time correlator of the $\bm n$-field can be expressed in terms of the
$\theta$ correlation function, $G(x-y) = \langle \theta(x) \theta(y)\rangle$:
\begin{equation}
  \label{eq:C_0}
C(r)=\langle n_\mu(r) n_\mu(0)\rangle= Re\langle e^{i\theta(r)} e^{-i\theta(0)}\rangle = e^{G(r)-G(0)} \ . 
\end{equation}
The $\theta$-field Green's function reads
\begin{eqnarray}
  \label{eq:corr_theta(r)}
  G(r)=\langle\theta (r) \theta (0) \rangle
  =\int\frac{d^2q}{(2\pi)^2}\frac{e^{i \bm{qr}}}{2\omega_{0q}} \ .
\end{eqnarray}
Hence
\begin{eqnarray}
  \label{Gr1}
  &&G(r)-G(0)
  =\int_0^{\Lambda}\frac{d^2q}{(2\pi)^2}\frac{e^{i \bm{qr}}-1}{2\omega_{0q}}=\\
  &&  \int_0^{\Lambda r}\frac{d\eta } {\eta}\int_0^{2\pi} 
  \frac{d\alpha[e^{i\eta\cos({\psi-\alpha)}} - 1]/(8\pi^2\sqrt{b_1}) }{ \sqrt{  (\cos^4{\alpha} + \sin^4{\alpha} ) + 2(b_2/ b_1) \sin^2{\alpha} \cos^2{\alpha}}} 
\nonumber
\end{eqnarray}
Here  the angle $\psi$  describes the orientation of the radius vector $\bm r=r(\cos{\psi},\sin{\psi})$ in the  2D plane $\{x,y\}$.
The first term in the square brackets is convergent at large $\eta$ while the
second term is logarithmically divergent. 
 Therefore, the first term just
provides the lower limit of integration, $\eta\sim 1$,   for the second one.
Hence, at $\Lambda r \gg 1$
\begin{eqnarray}
\label{Gr2}
  G(r)-G(0)
  \approx -\frac{\kappa_1}{4\pi\sqrt{b_1}}\int_1^{\Lambda r}\frac{d\eta}{\eta}   =-\zeta \ln(\Lambda r) \ ,
\end{eqnarray}
where
\begin{eqnarray}
  \label{zet}
&&  \zeta=\frac{\kappa_1}{4\pi \sqrt{b_1}}\\
  &&\kappa_1 = \int_0^{2\pi}
  \frac{d\alpha/(2\pi)}{ \sqrt{  (\cos^4{\alpha} + \sin^4{\alpha} ) + 2(b_2/ b_1) \sin^2{\alpha} \cos^2{\alpha}}} \nonumber
\end{eqnarray}
Thus the spin-spin correlator at the critical point, $\rho\to 0$, decays algebraically
\begin{equation}
  \label{eq:C_reg}
C(r) =  
\left(\frac{ 1 }{\Lambda r}\right)^{\zeta} \ .
\end{equation}
Interestingly, this correlator decay for the XY-Lifshitz spin liquid 
is similar to that in the dimer models at 
the Rokhsar-Kivelson critical point~\cite{Ardonne2004, Ghaemi2005}.

\subsection{Static magnetization}
The staggered magnetization in the Neel phase, $\rho > 0$,  is
\begin{eqnarray}
  \label{nz1}
  \langle n_x\rangle=Re \langle e^{i\theta}\rangle=e^{-(\langle \theta^2\rangle - \langle \theta\rangle^2)/2} \ .
\end{eqnarray}
We choose the $x$-axis as the direction of spontaneous symmetry breaking in the Neel phase. 
The  fluctuation of phase $\theta$ is given by Eq.(\ref{t2}).
According to Eq.(\ref{dispq}) there is a regime crossover at
\begin{eqnarray}
  \label{qm}
 q_{min} \sim \sqrt{\rho_r/b} \ .
\end{eqnarray}
At $q > q_{min}$ the dispersion is quadratic in q, and at $q < q_{min}$ the dispersion is linear in q.
Therefore, $q_{min}$ is the infrared cutoff in the logarithmically
divergent integral in (\ref{t2}). Hence
\begin{eqnarray}
  \label{nz2}
\langle \theta^2\rangle - \langle \theta\rangle^2=
\int_{qmin}^{\Lambda}\frac{d^2q}{(2\pi)^2}\frac{1}{2\omega_q}
=\zeta\ln(\Lambda/q_{min})  \ .
  \end{eqnarray}
This results in the following critical behaviour of the static magnetisation
\begin{eqnarray}\label{eq:beta_nz}
M\propto \langle n_x\rangle \propto \rho_r^{\zeta/4} \ .
\end{eqnarray}
It is easy to check that the static magnetisation in the spin-spiral phase
has the same scaling $ M\propto |\rho_r|^{\zeta/4}$.

Thus we come to the following conclusions of this section.\\
(i) The SL is realised only at the critical point, $\rho_r=0$.\\
(ii) The spin-spin correlator in the SL phase decays algebraically
$\propto r^{-\zeta}$.\\
(iii) The static magnetisation away from the SL point scales as 
$\rho_r^{\zeta/4}$. The critical index is not a universal number, but it depends
on parameters of the system, see Eq.(\ref{zet}). For the $J_1-J_3$ model
the critical point is $J_{3c}\approx 0.4$, see Fig.~\ref{fig:O2}c.
Hence using Eq.(\ref{parS}) we estimate  $\zeta/4 \approx 0.25$.\\
(iv) Because of the logarithmic correction in Eq.(\ref{rcor})
the critical scaling is not just a power, but there is a  logarithmic
dependence.\\
(v) The incommensurate wave vector scaling in the spin spiral phase also
has a logarithmic correction, Eq.(\ref{Qb1}).

\section{Finite temperature properties}\label{finiteT}
\subsection{Spin stiffness renormalization}
\begin{figure}[h!]
\includegraphics[scale=0.5]{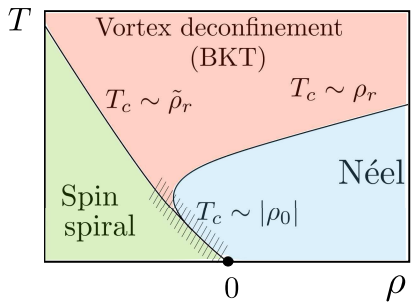}
\caption{Sketch of the $T-\rho$ phase diagrams of the Quantum Lifshitz transition in XY model for a finite system size $L$. 
   Neel and Spin spiral phases at $T>0$ possess a quasi-long range order and algebraic spin-spin correlations.}
\label{fig:phase_diagr}
\end{figure}

In this section we consider the effects of finite temperature T on  Lifshitz criticality.
 We assume that the temperature is much smaller than the
energy ultraviolet cutoff, $T \ll \sqrt{b}\Lambda^2$, but it can be larger
or comparable with $\rho$.
It is obvious that with these conditions temperature does not influence the
ultraviolet renormalization (\ref{dro}). However, the infrared renormalization
(\ref{eq:rho_r_sub1}) is changed. The $\frac{1}{\omega_q}$
term in the square brackets in (\ref{eq:rho_r_sub1}) should be replaced by a term containing a bosonic
occupation number factor $\frac{1}{\omega_q} \to \frac{1}{\omega_q}
(1+2n_q)$, where
\begin{eqnarray}
  \label{nq}
  n_q=\frac{1}{e^{\omega_q/T}-1} \ .
\end{eqnarray}
Hence Eq.(\ref{eq:rho_r_sub1}) modified for the case of finite $T$ reads
\begin{eqnarray}
  \label{eq:rho_r_subT}
&&\rho_r = \rho +  
 \left(3b_1+\frac{b_2}{2}\right)
  \int \frac{d^2 q}{(2\pi)^2}  q_x^2 \left[\frac{1}{\omega_q} -
    \frac{1}{\omega_{0q}} \right]\nonumber\\
  &&+
  \left(3b_1+\frac{b_2}{2}\right)\int
  \frac{d^2 q}{(2\pi)^2} q_x^2 \frac{2n_q}{\omega_q}\ .
\end{eqnarray}
This equation is written for the N\'eel phase.
For low temperature, $T \lesssim \rho$, the last line in
Eq.(\ref{eq:rho_r_subT})
is negligible and we return back to (\ref{eq:rho_r_sub1}).
However, for $\rho \ll T \ll \sqrt{b}\Lambda^2$ temperature is
significant and evaluation of (\ref{eq:rho_r_subT}) gives
\begin{equation}
  \label{eq:rho_rT}
  \rho_r = \frac{\rho + \frac{3T \kappa_2}{2\pi}\left(1+\frac{b_2}{6b_1}\right)
    \ln\left( \frac{\sqrt{b_1}T}{|\rho_r|}\right) }{1 + \frac{3\kappa_0}{8\pi \sqrt{b_1}}
\left(1+\frac{b_2}{6b_1}\right)
\ln\left(\frac{\Lambda^2 b_1}{|\rho_r|}\right) }\ ,
\end{equation}
where
\begin{eqnarray}
  \label{k2}
\kappa_2 = \int_0^{2\pi} \frac{d\alpha/(2\pi) \ \cos^2{\alpha}}
{ (\cos^4{\alpha} + \sin^4{\alpha} ) + 2(b_2/ b_1) \sin^2{\alpha} \cos^2{\alpha}} \ .
\end{eqnarray}
An equation similar to (\ref{eq:rho_rT})
was obtained in Ref.\cite{Ghaemi2005} for a particular case $\rho=0$.
Note that Eq. (\ref{eq:rho_rT}) has a positive solution, $\rho_r > 0$,
for small negative $\rho$. This means that the stability region of the
commensurate fluctuating ``Neel'' phase at finite temperatures
extends towards negative $\rho$, as shown in the phase diagram
in Fig.~\ref{fig:phase_diagr}.

\subsection{Magnon lifetime effect}
\begin{figure}[h!]
\includegraphics[scale=0.4]{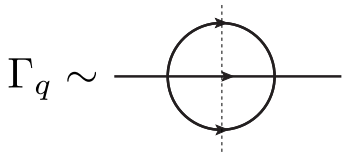}
\caption{Two-loop self-energy diagram that generates width of the  magnon. }
  \label{fsun}
\end{figure}

When considering Eq.(\ref{eq:rho_rT}) at very small $\rho_r$ we will face a serious problem.
The equation (\ref{eq:rho_rT}) does not have a
solution $\rho_r=0$. Hence the transition line in Fig.~\ref{fig:phase_diagr}
between the fluctuating N\'eel state and the fluctuating spin spiral state is
undetermined.
To resolve this problem we need to consider  perturbative and
non-perturbative (topological)  fluctuations. In this subsection
we consider perturbative fluctuations.
Equation (\ref{eq:rho_rT}) is derived in the single loop approximation, see
Fig. \ref{fig:loop}. In this approximation the self-energy does not have an
imaginary part. The imaginary part arises due to the double loop diagram
shown in  Fig. \ref{fsun}.
The four-magnon vertex in Fig.\ref{fsun} 
is generated by quartic terms  $b(\partial \theta)^4$ in
Eq.(\ref{eq:L_full_theta}).
  In what follows we denote by $q$ the momentum of the probe magnon, and $k_1$,
  $k_2$, $k_3$ are momenta of magnons from the heat bath.
  There are three types of the scattering processes contributing to the
  width:
  (i) Decay: $q = k_1+k_2+k_3$,
  (ii) Raman: $q+k_1 = k_2+k_3$,
  (iii) Fusion:  $q+k_1+k_2 = k_3$  
  We consider the most important  Raman process, the corresponding width reads (see e.g. Ref. \cite{Scammell2017}):
\begin{eqnarray}
  \label{eq:Gamma_q}
  \Gamma_q &\sim& \frac{(1-e^{-\omega_q/T})}{\omega_q} b^2 q^2 \int \frac{d^2 k_1}{2\omega_1} \frac{d^2 k_2}{2\omega_2} \frac{d^2 k_3}{2\omega_3} k_1^2 k_2^2 k_3^2
\nonumber \\
 & \times&  
n_1 (1+n_2)(1+n_3)\, \delta(\omega_q+\omega_1-\omega_2-\omega_3)\nonumber\\
&\times&  \delta^{(2)}({\bm q} + {\bm k}_1 - {\bm k}_2 - {\bm k}_3) \ .
\end{eqnarray}
We are interested in the case $\rho_r=0$, hence the dispersion is
$\omega_q = \sqrt{b}q^2 \ll T$,
$\omega_i =\sqrt{b}k_i^2 \ll T$.
The occupation numbers can be replaced as $n_i \to \frac{T}{\omega_i}$.
Evaluation of the integral in (\ref{eq:Gamma_q}) is straightforward, the
result is
\begin{eqnarray}
  \label{gq}
  \Gamma_q \sim \frac{T^2}{\sqrt{b}}\frac{q^2}{\omega_q^2} \ .
\end{eqnarray}
This estimate is valid if $\Gamma_q \ll \omega_q$ .
At very small $q\to0$ this condition is violated and hence in the right hand side
of (\ref{gq}) we have to replace $\omega_q \to \Gamma_q$.
This immediately gives the following estimate for $\Gamma_q$ at very
small $q$
\begin{eqnarray}
  \label{gqs}
  \Gamma_q \sim \frac{T^{2/3}}{b^{1/6}}q^{2/3} \ .
  \end{eqnarray}
This width results in the effective infrared cutoff in the $q-$integration
in the last line of Eq.(\ref{eq:rho_r_subT}). Hence, with account
of the width, Eq.(\ref{eq:rho_rT}) is replaced by
\begin{eqnarray}
  \label{eq:rho_rTT}
  \rho_r &=& \frac{\rho + \frac{3T \kappa_2}{2\pi}
    \ln\sqrt{b}}
      {1 + \frac{3\kappa_0}{8\pi \sqrt{b_1}}
\left(1+\frac{b_2}{6b_1}\right)
\ln\left(\frac{\Lambda^2 b_1}{|\rho_r|}\right) }\nonumber\\
      &\to&
      \frac{\rho + {\cal C} T}
      {1 + \frac{3\kappa_0}{8\pi \sqrt{b_1}}
\left(1+\frac{b_2}{6b_1}\right)
\ln\left(\frac{\Lambda^2 b_1}{|\rho_r|}\right) }      \ .
\end{eqnarray}
The $q-$integral in the numerator of this equation has been calculated
with logarithmic accuracy, $\ln(q_T/q_{min})\gg 1$, where $q_{min}$ is the
lifetime infrared cutoff that follows from  Eq.(\ref{gqs}).
However, in the end, the logarithm proved to be not large.
Therefore, we must replace it by a constant ${\cal C}$ that we cannot calculate
within accuracy of the method. We can only claim that the constant is positive
since the integral in the second line of Eq.(\ref{eq:rho_r_subT}) is positive.
The transition line in Fig.~\ref{fig:phase_diagr} between the fluctuating
N\'eel state and the fluctuating spin spiral state is given
by the condition $\rho_r=0$, that results in the condition
\begin{eqnarray}
  \label{trans}
  \rho =- {\cal C} T.
\end{eqnarray}

\subsection{Finite temperature spin-spin correlators}
  Let us consider the spin-spin correlator
at the transition line $\rho_r=0$, see  Fig.~\ref{fig:phase_diagr}.
At a finite temperature Eq.(\ref{Gr1}) is transformed to
\begin{eqnarray}
  \label{Gr1T}
&&  G(r)-G(0)
  =\int_0^{\Lambda}\frac{d^2q}{(2\pi)^2}\frac{e^{i \bm{qr}}-1}{2\omega_{0q}}
  (1+2n_q)
  \nonumber\\
&&\approx T\int_0^{q_T}\frac{d^2q}{(2\pi)^2}\frac{e^{i \bm{qr}}-1}{\omega_{0q}^2}
+\int_{q_T}^{\Lambda}\frac{d^2q}{(2\pi)^2}\frac{e^{i \bm{qr}}-1}{2\omega_{0q}}
\end{eqnarray}
where $q_T = \sqrt{T}/b^{1/4}$. The first integral in this equation is 
infrared logarithmic divergent. As was discussed in the previous subsection
the integral has to be regularised by the finite width, $\omega_{0q}^2 \to
\omega_{0q}^2+\Gamma_q^2$.
Evaluation of the integrals in (\ref{Gr1T}) is straightforward.
At $q_Tr \ll 1$ the correlator $C(r)$ decays with distance exactly like that
in the zero temperature SL, Eq.(\ref{eq:C_reg}), and at $q_Tr \gg 1$ it
very quickly decays to zero.

Away from the critical line, $\rho_r \ne 0$,  at sufficiently large distances,
$r\gg \sqrt{b/|\rho_r|}$, the spin-spin correlator decays according to
the standard XY-model algebraic law 
\begin{eqnarray}
C(r) \propto \frac{1}{r^{T/(2\pi|\rho_r|)}}\ .
\end{eqnarray}

\subsection{The role of vortices}
A single vortex in the XY-model has energy
\begin{eqnarray}
  \label{evort}
  E_{vortex}= E_{core}+\pi\rho\ln(L/a) \ ,
\end{eqnarray}
where $L$ is the size of the vortex (size of the sample), and
$E_{core}$ is the energy of the vortex core.
Hence at a finite temperature the  free energy per vortex reads
\begin{equation}
  \label{eq:F_vortex}
F_{vortex} = E_{core} + \pi \rho \ln\left(\frac{L}{a}\right) - T \ln\left(\frac{L^2}{a^2}\right) \ .
\end{equation}
The third term in this equation is due to the entropy of the vortex gas
$S_v\sim\ln\left(\frac{L^2}{a^2}\right)$.
In the usual unfrustrated XY-model ($J_3=0$), the core energy $E_{core} \sim \rho \sim J_1$.
Therefore the core energy does not play a significant role.
In the limit $L\to \infty$  the vortex proliferation is energetically
favourable ($F_{vortex}<0$) at the  temperatures above the BKT temperature
$T>T_{BKT} = \frac{\pi}{2}\rho$. 
  Formally near  the Lifshitz point the situation is similar and
  we can write the following equation for $T_{BKT}$.
  \begin{eqnarray}
    \label{bkt}
T_{BKT}=\frac{\pi}{2}|\rho| \ .
\end{eqnarray}
  In particular $T_{BKT}=0$ at $\rho=0$.
  Obviously Eq.(\ref{bkt}) is not consistent with Fig.\ref{fig:phase_diagr}.
The explanation is that this equation represent a rather formal statement.
While at the Lifshitz point the spin stiffness is approaching zero due to
frustration, $\rho \to 0$,
the core energy remains finite, $E_{core} \sim J_1$, see Ref.\cite{Kharkov17}
According to Eq.(\ref{eq:F_vortex}) in the limit $L\to \infty$ and at
$\rho = 0$  vortexes proliferate at any nonzero temperature.
This is consistent with Eq.(\ref{bkt}).
However, to make this point valid  the entropy in Eq.(\ref{eq:F_vortex})
must dominate over the core energy, hence the
sample size must be sufficiently large, $L > a \times e^{E_{core}/2T}$.
This can be a huge number, thousands, millions, or billions of lattice spacings
that is never achieved in a real sample or in numerical experiments on a finite cluster.

Thus formally mathematically the transition line in Fig.~\ref{fig:phase_diagr}
originating from $\rho=0$ consists of two diverging phase boundary lines that are exponentially close to each other. The intermediate phase between the lines is  the deconfined vortex phase.
However, in a real sample, say $1000 \times 1000$ sites, still there is 
single transition line that at a certain temperature starts to diverge into
two BKT lines as it is shown in Fig.~\ref{fig:phase_diagr}.
The position of the divergence point depends on the sample size dramatically,
so the notion of the phase diagram near the divergence point is poorly
defined.

To summarise this subsection, at finite system size and at a
sufficiently low temperature the $O(2)$ topological excitations (vortices) are
irrelevant near the Lifshitz point.  This is because the vortex core energy
remains finite even if the spin stiffness is zero.
Interestingly, in the $O(3)$ case (isotropic Heisenberg model) the energy
of the topological excitation, skyrmion,  scales proportionally to the spin stiffness.
Therefore, in this case topological excitations are relevant near the Lifshitz point.\\

\section{Discussion and conclusion.}\label{sec:concl}
  We have considered the  Lifshitz quantum phase transition  problem for the 2D
  frustrated XY-model. Here are the conclusions.\\
{\bf  (i)} We have performed numerical series  expansion calculations for $J_1-J_3$
  model on the square lattice ($S=1/2$). The calculations indicate that
  the Lifshitz point behaviour  is very different for the
  XY- and for the SU(2)-symmetric versions of the model.\\
{\bf  (ii)} Motivated by the numerics we performed field theory
  analysis of the XY-Lifshitz criticality. This analysis results in
  the following points.\\
{\bf (iii)} The Lifshitz spin liquid phase exists only at the Lifshitz point.
  This is different from the SU(2) case where the spin liquid phase occupies a
  finite interval in the  parameter space.\\
{\bf (iv)}  At zero temperature we calculate nonuniversal  critical exponents in the
  N\'eel and in the spin spiral state and relate them to properties of the spin
  liquid.\\
{\bf (v)}   We also solve the transition problem  at a finite temperature,
  calculate the critical exponents, and discuss the role of the magnon
  lifetime on the finite temperature critical behaviour.\\
{\bf  (vi)} We  show that the topological excitations are irrelevant at low
  temperature even near the critical point.

\begin{acknowledgments}
  We thank Matthew O'Brien for discussions
  and Anders Sandvik for important communications.
  The work has been supported by the Australian Research Council
  No DP160103630.    
\end{acknowledgments}


\newpage
\appendix

\end{document}